\documentclass[a4paper,UKenglish,cleveref, autoref, thm-restate]{lipics-v2021}
\hideLIPIcs  %uncomment to remove references to LIPIcs series (logo, DOI, ...), e.g. when preparing a pre-final version to be uploaded to arXiv or another public repository

\usepackage{myacronym}
\newacro{ENAC}{École Nationale de l'Aviation Civile}
\newacro{LTL}{Linear Temporal Logic}
\newacro{pLTL}{Past Linear Temporal Logic}
\newacro{SMT}{Satisfiability Modulo Theories}
\newacro{BMC}{Bounded Model Checking}
\newacro{CPGE}{Classe Préparatoire aux Grandes Ecoles}
\newacro{ISI}{Ingénierie des Systèmes Interactifs}
\newacro{IENAC APPR}{Ingénieur ENAC par apprentissage}
\newacro{SI}{Systèmes d'Information}
\newacro{IATSED}{Master in International Air Transport System Engineering and Design}
\newacro{VV}[V\&V]{Verification \& Validation}
\newacro{PDR}{Property Directed Reachability}
\newacro{DOM}{Document Object Model}
\newacro{JIT}{Just-In-Time}
\newacro{VCS}{Version Control System}
\newacro{CRDT}{Conflict-Free Replicated Data Type}
\newacro{GenAI}{Generative Artificial Intelligence}
\newacro{LLM}{Large Language Model}

\usepackage{tikz}

\usepackage{lustre}
\usepackage{float}
\newfloat{lstfloat}{htbp}{lop}
\floatname{lstfloat}{Listing}
\Crefname{lstfloat}{Listing}{Listings}
\crefname{lstfloat}{listing}{listings}

\usepackage{amssymb}
\usepackage{pifont}
\newcommand{\cmark}{{\color{green!80!black}\checkmark{}}}
\newcommand{\xmark}{{\color{red}\ding{55}}}
\newcommand{\smark}{\ensuremath{{\color{orange}\thicksim}}}

\title{Teaching Synchronous Dataflow Modelling with Learn-Heptagon}

\author{Pierre-Loïc Garoche}{Fédération ENAC ISAE-SUPAERO ONERA, Université de Toulouse, France}{pierre-loic.garoche@enac.fr}{https://orcid.org/0000-0002-0513-6076}{}
\author{Basile Pesin}{Fédération ENAC ISAE-SUPAERO ONERA, Université de Toulouse, France}{basile.pesin@enac.fr}{https://orcid.org/0000-0002-3575-7770}{}

\authorrunning{Pierre-Loïc Garoche and Basile Pesin} % mandatory. First: Use abbreviated first/middle names. Second (only in severe cases): Use first author plus 'et al.'

\Copyright{Pierre-Loïc Garoche and Basile Pesin} % mandatory, please use full first names. LIPIcs license is "CC-BY";  http://creativecommons.org/licenses/by/3.0/


 % mandatory; please add comma-separated list of keywords

\category{} %optional, e.g. invited paper

\relatedversion{} %optional, e.g. full version hosted on arXiv, HAL, or other respository/website
\acknowledgements{We thank Aurélien Anglade for his significant contributions to the development of Learn-Heptagon.}

\nolinenumbers %uncomment to disable line numbering

\EventEditors{Shriram Krishnamurthi and Thomas Zeume}
\EventNoEds{2}
\EventLongTitle{TEAL 2026: Tools for Educational Activities in Logic}
\EventShortTitle{TEAL 2026}
\EventAcronym{CVIT}
\EventYear{2026}
\EventDate{July 25, 2026}
\EventLocation{Lisbon,Portugal}
\EventLogo{}
\begin{document}

\maketitle

\begin{abstract}
  Lustre is a synchronous dataflow language designed to implement safety-critical embedded software.
In addition to writing executable programs, the language doubles as a program logic, used for writing specification as synchronous observers or assume-guarantee contracts that specify properties of these programs.
These specifications may be used during testing or proved exhaustively using model-checking tools.
We taught a course on Lustre to last year engineering students.
To streamline the learning experience and avoid technical issues, we developped an online application, Learn-Heptagon, which allows for writing, simulating, and proving properties of Lustre programs.
This paper presents the application and the associated lesson plan.

\end{abstract}

\section{Introduction}

%% In this first section, we describe the Lustre language, along with its feature
%% and ecosystem, focusing particularly on the formal verification features.
%% Then we introduce the profile of the engineering students we taught Lustre to,
%% and justify why such a course is a worthwhile addition to their syllabus. 

Embedded systems are traditionally programmed using low-level languages such as
C.
However, programming with these languages is error-prone, due to potential
memory manipulation errors, and concurrency pitfalls.
It is also difficult to trace high-level system requirements to a low-level implementation.
Therefore, these languages are not well suited to programming safety-critical
embedded systems.
Instead, new programming paradigms and dedicated languages have been proposed.
In this paper, we describe an interactive online application for teaching such a language.

\subsection{Synchronous Programming and the Lustre Language}

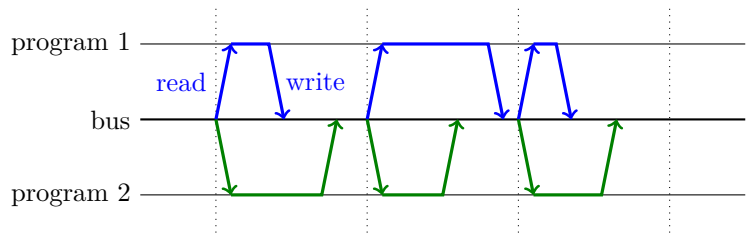
\begin{figure}
  \newcommand{\executiontop}[2]{
  \begin{scope}[very thick,blue,->]
  \draw (#1,0) -- ++(.2,1);
  \draw (#1+.2,1) -- ++(#2,0) -- ++(.2,-1);
  \end{scope}
}
\newcommand{\executionbot}[2]{
  \begin{scope}[very thick,green!50!black,->]
  \draw (#1,0) -- ++(.2,-1);
  \draw (#1+.2,-1) -- ++(#2,0) -- ++(.2,1);
  \end{scope}
}

\begin{center}
\begin{tikzpicture}
  \draw[thick] (0,0) -- ++(8,0) node[pos=0,anchor=east] {bus};
  \draw[] (0,1) -- ++(8,0) node[pos=0,anchor=east] {program 1};
  \draw[] (0,-1) -- ++(8,0) node[pos=0,anchor=east] {program 2};
  \foreach \x in {0,...,3} \draw[dotted] (1+2*\x,-1.5) -- (1+2*\x,1.5);

  \begin{scope}[blue]
    \node[anchor=east] at (1,.5) {read};
    \node[anchor=west] at (1.8,.5) {write};
  \end{scope}
  \executiontop{1}{.5};
  \executionbot{1}{1.2};

  \executiontop{3}{1.4};
  \executionbot{3}{.8};
  
  \executiontop{5}{.3};
  \executionbot{5}{.9};
\end{tikzpicture}
\end{center}
  \caption{Example execution of two programs under the synchronous hypothesis}
  \label{fig:synchronous-hypothesis}
\end{figure}

Synchronous programming~\cite{BenvenisteBer:SyncApp:1991} is a paradigm where 
physical time is abstracted by discrete, logical time steps.
Parallel components of a program communicate synchronously at these time steps,
which usually occur at a set physical period.
This abstraction makes composition of programs trivial, and 
remains valid as long as one execution of the program ``fits'' inside the
physical period.
This is the so-called ``synchronous hypothesis'', sketched in
\cref{fig:synchronous-hypothesis}: both programs read their inputs at the start
of the period, symbolised by the dotted lines.
They write their outputs as soon as they are finished.
The order in which they write does not matter as long
as they do before the start of the next period.

Lustre~\cite{CaspiEtAl:Lustre:1987} is a synchronous dataflow programming
language.
Lustre programs, such as the one presented in \cref{lst:counter} are composed of
nodes which are synchronous functions.
Each node receives inputs: for \texttt{sum}, an integer input
\texttt{i} and a boolean input \texttt{r}.
It computes outputs: for \texttt{sum}, a single integer output
\texttt{s}.
A node may also have internal variables, such as \texttt{ps} for sum.
Each variable represents an infinite stream of values of the specified type.
The chronogram on the right represent one possible execution of node
\texttt{sum}: at each time step, the node receives inputs \texttt{i} and
\texttt{r}, and computes the values of \texttt{s} and \texttt{ps} according to
the equations between \texttt{let} and \texttt{tel}.
This particular node computes the cumulative sum of \texttt{i},
resetting every time \texttt{r} is true.
The first equation specifies the value of internal variable \texttt{ps}.
At the first step (left of the initialisation arrow \lus{->}) it is 0.
At all other steps, it is equal to 0 if \lus{r} is true.
Otherwise, it is equal to the previous value of \lus{s} (\lus{pre s}).
Output \lus{s} is simply equal to the sum of \lus{ps} and \lus{i}.
The history on the right of \cref{lst:counter} shows a sample execution of node \lus{sum}, with specified inputs \lus{i} and \lus{r}.
The \lus{counter} node instantiates \lus{sum} with a fixed parameter for
\lus{i}.
More generally, a node may call any number of previously defined nodes, but
recursive calls are not allowed.

\begin{lstfloat}
  \begin{minipage}{.49\textwidth}
  \begin{lstlisting}
node sum(i:int; r:bool) returns (s:int)
var ps:int;
let
  ps = 0 -> if r then 0 else (pre s);
  s = i + ps;
tel

node counter(r:bool) returns (c:int)
let c = sum(1, r); tel
  \end{lstlisting}
  \end{minipage}
  \hfill
  \begin{minipage}{.47\textwidth}
    \begin{tabular}{c|lllllll}
      \lus{i} & 1 & 3 & 2 & 2 & 1 & 4 & \ldots \\
      \lus{r} & F & F & F & T & F & F & \ldots \\
      \hline
      \lus{pre s} &  & 1 & 4 & 6 & 2 & 3 & \ldots \\
      \lus{ps} & 0 & 1 & 4 & 0 & 2 & 3 & \ldots \\
      \hline
      \lus{s} & 1 & 4 & 6 & 2 & 3 & 7 & \ldots
    \end{tabular}
  \end{minipage}
  \caption{The \texttt{sum} and \texttt{counter} nodes, with an example
    execution of \texttt{sum}}
  \label{lst:counter}
\end{lstfloat}

Lustre programs are compiled efficiently to C code~\cite{BiernackiEtAl:ClockDirected:2008}.
Therefore, a trusted compiler guarantees tracability from the high-level model of the system as a
Lustre program down to the low-level imperative code.
This tracability is mandatory for safety-critical embedded systems.
To this end, Scade~\cite{Ansys:Scade}, a compiler for a Lustre-like language, was
qualified under DO-331~\cite{DO-331} which means it can be used in safety-critical
embedded systems like avionics or nuclear power plants.
%% The latest version, Scade~6~\cite{ColacoPagPou:SCADE:2017}, incorporates several
%% extends of Lustre: hierarchical state
%% machines~\cite{ColacoPagPou:StateMachines:2005}, higher-order
%% functions~\cite{CaspiHamPou:LucidSynchrone:2008}, hybrid
%% modelling~\cite{BenvenisteEtAl:EMSOFT:2011}, etc.
%% the basis of SCADE~Suite, a model-based development environment used
%% for implementing safety-critical embedded software, such as avionics systems,
%% electric vehicle control software, etc.
%% The graphical representation of SCADE~Suite %% , shown in \cref{fig:scade},
%% is based
%% on block diagrams and translated directly to a Scade~6 dataflow synchronous
%% program for compilation.

%% \begin{figure}
%%   \begin{center}
%%   \includegraphics[width=\textwidth]{figures/scade.png}
%%   \end{center}
%%   \caption{SCADE Suite: an industrial tool based on Lustre}
%%   \label{fig:scade}
%% \end{figure}

\subsection{Model Checking of Lustre Programs}

One of the main advantage of the Lustre language is that its design facilitates
formal verification of safety properties (i.e. invariant properties) of programs.
To do so, these properties may be expressed as invariants of reachable states.
Logical formulas are implemented by boolean Lustre expressions, which means the
same language is used for both writing programs and their properties.
Properties are implemented by so-called ``synchronous observers''~\cite{DieumegardEtAl:SyncObs:2015}: nodes that
observe the inputs and outputs of other nodes.
%% \acp{LTL}~\cite{Pnueli:LTL:1977} logical formulas, implemented by boolean Lustre expressions.

For instance, the node \lus{observe_sum} presented in
\cref{lst:sofar-contract} at left checks that, if input \texttt{i} of \texttt{sum} has
always been non-negative, then output \texttt{s} is non-negative.
The output, \texttt{ok}, should always be true.
It is defined using the \texttt{sofar} node, which implements the
\ac{pLTL}~\cite{Lichtenstein:PLTL:1985} operator $\boxminus$ (always true in the past).
%% \texttt{sofar} node presented in \cref{lst:sofar-contract} at left: as long as its input
%% \texttt{b} is true, it will stay true, but will switch irremediably to false as
%% soon as \texttt{b} is false. Therefore, at each time step, \texttt{s} is
%% true if and only if \texttt{b} has always been true in the past.

Modern versions of Lustre have a syntax to specify properties of nodes in
contracts.
A contract equivalent to the observer described above is presented in
\cref{lst:sofar-contract} at right.
It follows the syntax of the Kind2 model checker~\cite{ChampionEtAl:Kind2:2016}.
A contract \luskw{assume}s pre-conditions on the inputs of the node, and
\luskw{guarantee}s a property on the inputs and outputs.
%% In the case of the \texttt{sum} node, if we assume that the input is never
%% negative, then the cumulative sum should never be negative.
Note that the semantics of $\luskw{assume}\ e$ is ``$e$ has always been true in
the past''. This contract is indeed equivalent to using the \texttt{sofar} node.
%% an equivalent contract without an \luskw{assume} clause would be
%% \lus{guarantee sofar(i >= 0) => (s >= 0)}.
%% Since the guarantee must be true at every step, this is also equivalent to
%% \lus{guarantee sofar(i >= 0) => sofar(s >= 0)}.

\begin{lstfloat}
  \begin{minipage}{.45\textwidth}
    \begin{lstlisting}
node sofar(b:bool) returns (s:bool)
let s = b -> b and (pre s); tel

node observe_sum(i:int; r:bool; s:int)
returns (ok:bool)
let ok = sofar(i >= 0) => (s >= 0) tel
    \end{lstlisting}
  \end{minipage}
  \hfill
  \begin{minipage}{.49\textwidth}
    \begin{lstlisting}
node sum(i:int; r:bool) returns (s:int)
(*@contract
  assume i >= 0;
  guarantee s >= 0; *)
...
    \end{lstlisting}
  \end{minipage}
  \caption{Synchronous observer (left) and contract (right) for the \texttt{sum} node}
  \label{lst:sofar-contract}
\end{lstfloat}

Synchronous observers and contracts may be used for testing purposes: while
running the node, the contract is evaluated at each step on the actual inputs and
outputs of the node to ensure it is not violated.
However, ensuring that contract violations never occurs requires
good test coverage of the code, which may require searching for specific
inputs~\cite{Papailiopoulou:Coverage:2011}.

Another interesting approach is using model-checking to formally prove that contracts are never violated.
Several model-checkers dedicated to the Lustre language have been introduced~\cite{HagenTin:Kind:2008}.
The simplest ones rely on $k$-induction~\cite{SheeranSinSta:kInduction:2000}.
If the goal is to prove that a property $P$ is true at each time step, then the solver proves that:
\begin{itemize}
\item $P$ is true at the first $k$ time steps
\item if $P$ is true for time steps $n,\ldots{},n+k$, then $P$ is true for time step $n+k+1$
\end{itemize}
For example, the property for the \texttt{sum} node is $k$-inductive with $k=1$.
Note that this is only true when we assume idealized mathematical integer;
using machine integers, \texttt{s} may overflow back to a negative value.
This assumption is often made by model-checkers.

More recent model-checkers, such as Kind~2~\cite{ChampionEtAl:Kind2:2016},
encode the semantics of the program as a transition system.
Then, they use several algorithms, such as k-induction or
\ac{PDR}~\cite{Bradley:IC3:2011} to prove that the property of interest is
an invariant.
Elementary subgoals of the proof are discharged to \ac{SMT} solvers.
%% combine k-induction with \ac{BMC}, which
%% represents the program as a transition system.
These more complex algorithms allow, in particular, to prove some non
$k$-inductive properties by strengthening them with new invariants.

\subsection{Teaching Synchronous Dataflow Modelling}

%% In our experience, courses that use dedicated modelling or programming languages
%% often suffer from the lack of user-friendly tooling.
In the Lustre ecosystem, the most likely candidate languages to be used as a support for a
course on dataflow modelling are the industrial language Scade~6, and the
academic languages Lustre v4/v6, Heptagon, LustreC, and Kind 2.
The most obvious candidate would be Scade~6 itself. However, Scade~6 is
closed-source and licences are relatively difficult to obtain in the academic
setting.
Moreover, its block-diagram based user interface is, in our opinion, quite
daunting, and detracts from the simplicity of the underlying language.
On the other hand, the academic languages are all free but all support a
different set of features.
In \cref{tbl:dialects}, we consider the support for model-checking and for the
two most important extensions introduced in Scade~6: 
state machines and arrays.
Only Scade~6, Heptagon and LustreC support state machines.
Additionally, Lustre~v4, LustreC and Kind~2 only support limited features of arrays.
In our experience, using two different languages (e.g. Heptagon and Kind 2) in the
same course to illustrate the different features is not a satisfying solution:
students get confused by the (small) discrepencies in syntax between the two languages,
which impedes learning the concepts common to the two languages.

\begin{table}
  \begin{center}
\begin{tabular}{l|ccccc}
  & Arrays & State Machines & Model Checking & Free Software & Online Interface \\
  \hline
  Scade~6~\cite{Scade6Language:2016} & \cmark{} & \cmark{} &
  \cmark{} & \xmark{} & \xmark{} \\
  Lustre v4~\cite{Raymond:Lustre4:1992} & \smark{} & \xmark{} & \cmark{} &
  \cmark{} & \xmark{} \\
  Lustre v6~\cite{JahierRayHal:Lustre6:2019} & \cmark{} & \xmark{} &
  \xmark{} & \cmark{} & \xmark{} \\
  Heptagon~\cite{Heptagon-Tutorial:2017} & \cmark{} & \cmark{} & \xmark{} &
  \cmark{} & \xmark{} \\
  LustreC~\cite{Garoche:lustrec:2012} & \smark{} & \cmark{} & \cmark{} & \cmark{} & \xmark{} \\
  Kind 2~\cite{ChampionEtAl:Kind2:2016} & \smark{} & \xmark{} & \cmark{} &
  \cmark{} & \cmark{} \\
\end{tabular}
\end{center}
\caption{Most common Lustre dialects used in teaching. For each dialect and feature, \cmark{} means the feature is fully supported, \smark{} means it is partially supported and \xmark{} means it is not supported.}
\label{tbl:dialects}
\end{table}

Moreover, the tools for academic languages (except for Kind 2) are usually
command-line based, which students may not be comfortable with.
This increases the extraneous cognitive load of students to the detriment of
learning the language.
Installing these tools may also require complex installation procedure, which
might waste a substantial amount of time in relatively short modules, and in
some cases not work at all on students or faculty computers.

\subsection{Contributions and Outline}

To alleviate the issues described above, we developed a new application for
teaching Lustre modelling and programming, called Learn-Heptagon.
As its name implies, it is based on the Heptagon compiler, but also supports
model-checking of programs by translating them to Kind 2.
\Cref{sec:learn-heptagon} describes the design and implementation of this application.
We already used this application to teach two classes of students.
\Cref{sec:lesson-plan} discusses our lesson plan, and reports on our
experience teaching with this application.

\section{The Learn-Heptagon Application}
\label{sec:learn-heptagon}

In this section, we present our application, Learn-Heptagon, which is available at
\begin{center}
\url{https://learn-heptagon.vertmo.org/}
\end{center}
We first present and justify the high-level user needs that informed the
design of the application.
Then, we discuss its user interface design.
Finally, we describe some of the technical choices made for its implementation.

\subsection{User Needs and Software Requirements}

To define the requirements on the design of our application, we mostly refered to the
needs that had arised when teaching previous iterations of the course.
The results of this elicitation process is detailed below, with the resulting
software requirements in bold.

\newcommand{\req}[1]{\textbf{#1}}

In order to prevent back-and-forth between the computer and the exercise sheet,
it would be desirable for the application itself to
\req{display the specification of the exercise}; more specifically, it should
\req{display the exercise-specific instructions}.
To guide the students, we also provide the signature
of the expected node; in some cases, we also provide some pre-defined nodes.
The application should also
\req{automatically load and display the provided code canvas}.

Then, the application should
\req{allow the student to propose a solution for the exercise}.
In particular, the application should \req{support the full Lustre language}, along
with the main extensions supported in Scade~6: \req{support programs with state machines},
\req{support programs with arrays}.
Some exercises may also ask the student to reuse (e.g. call) a node defined in a
previous exercise.
Therefore, the application should also \req{support inter-exercise node calls}.

Of course, the application should \req{provide useful feedback to students}.
Since Lustre is a statically-typed language, it should in particular
\req{display syntax and typing error}.
Then, the application should \req{let the student test their solution}.
A test is run on a specific set of inputs.
For simple nodes, it is more practical to \req{let the student specify the testing input manually}.
For more complex nodes (e.g. modelling the full controller for an embedded 
system), entering a realistic trace of input lengthy enough might be
infeasible.
To cover these cases, the application should
\req{support several simulations dedicated to specific exercises}.
Finally, while testing allows the student to experiment with their solution, the
program should also allow them to
\req{automatically check if the solution to an exercise is correct}.
After an exam, the application should also
\req{allow the teacher to access the student's work} so that it may be graded.

Since verification of program properties is a focus of our course, the application
should also \req{support assume-guarantee contracts}.
The application should be able to
\req{model-check the properties expressed in contracts}, and
\req{display the result of model-checking}.
In particular, if the property does not hold, it should
\req{display potential counter-examples} to guide the student in refining their
code or contracts.

As explained in the context, one of the main issues encountered in previous
courses was to install the required software on the faculty/students machine.
To avoid this issue, the application should
\req{be usable without any installation process}. 
Since we cannot control which computers the student use, it should also
\req{run on any reasonably modern computer} and
\req{not depend on a specific operating system}.
Since students also tend to move between different computers, the application should
\req{enable students to access their work from several computers} directly.
For archiving purposes, it should also allow them to
\req{export and import their work in progress to and from files}.

Finally, we intend the application to be reusable for future lesson plans.
Therefore, the application should
\req{allow the teacher to easily define and modify lesson plans}.

%% \begin{itemize}
%% \item allow the student to read the specification of the exercise
%%   \begin{itemize}
%%   \item textual spec
%%   \item prototype of node
%%   \end{itemize}
%% \item allow the student to propose a solution to the exercise
%%   \begin{itemize}
  %% \item using arrays
  %% \item using using state machines
  %% \item using assume-guarantee contracts
  %% \item that depend on previous exercises
  %% \end{itemize}
%% \item inform the student of problems (typing, syntax) in their solution
%% \item allow the student to simulate their solution on a set of inputs
%%   \begin{itemize}
%%   \item specified manually
%%   \item specified systematically (by an expression)
%%   \item specified systematically by a simulator
%%   \end{itemize}
%% \item allow the student to check the correctness their solution
%% \item allow the student to check the validity of their contract
  %% \begin{itemize}
  %% \item display counterexample if found
  %% \end{itemize}
%% \item allow the student to import/export their work to readable 
%% \item allow the student to access their work from several computers
%% \item allow the teacher to easily modify lesson plans
%% \item allow the teacher to divide the lesson plan into several sessions
%% \item allow the teacher to access student's solutions to exams
%% \end{itemize}

\subsection{User Interface Design}

\begin{figure}
  \begin{minipage}{.65\textwidth}
  \includegraphics[width=\textwidth]{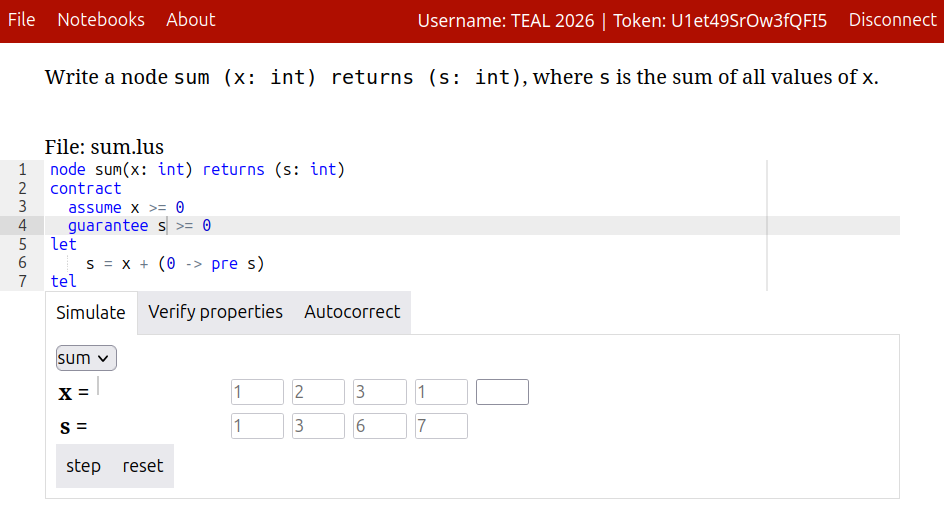}
  \end{minipage}
  \begin{minipage}{.34\textwidth}
    \begin{lstlisting}
node stopwatch(oo,rs,fr:bool)
returns (time:int)
    \end{lstlisting}
    \vspace{-1em}
    \begin{center}
  \includegraphics[width=.8\textwidth]{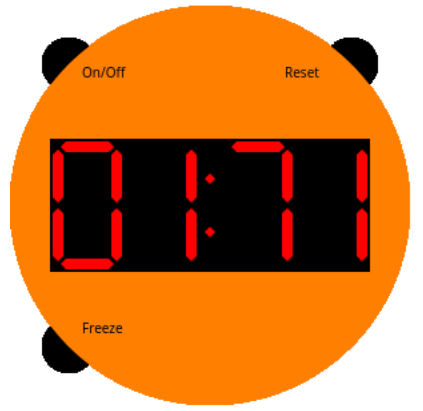}
  \end{center}
  \end{minipage}
  \caption{Screenshot of the Learn-Heptagon app (left) and the stopwatch 
    simulator (right)}
  \label{fig:screenshot1}
\end{figure}

To ensure the application is easily usable on any machine, we developped it as a web
app, which should run on any modern browser.
We now detail the different features of this app, shown on
\cref{fig:screenshot1}.

\paragraph*{Creating an account and login in}

On first connection, users are greeted with a login screen.
The app uses a lightweight authentication system based on tokens: on account
creation, users generate a token which they should copy.
They can then use it to reconnect to their account and retrieve their work.
The connection to the current account is maintained unless the user manually
disconnects (or clears the browser's cache).
This system is inspired by Learn~OCaml~\cite{Canou:LearnOcaml:2016}, an exercise platform
for the OCaml programming language~\cite{LeroyEtAl:OCamlMan:2025}.

\paragraph*{Notebooks}

The content of the application is structured into ``notebooks'', choosen using the top menu.
Each notebook corresponds to a sequence of exercises, which could correspond to
a full course, or one lesson.
A notebook is structured as a sequence of blocks of texts and code editors
which the students may edit; \cref{fig:screenshot1} shows a single (short) block
of text and a single code editor.
This structure is loosely based on that of Jupyter
Notebooks~\cite{JupyterNotebook}, an interactive environment for the Python language.

\paragraph*{Editing and feedback}

The application supports the full Heptagon~\cite{Heptagon-Tutorial:2017} variant
of Lustre.
Each code editor provides syntax highlighting for this language; to do se, we
use the Ace~\cite{AceEditor}, a web-based extensible code editor.
The program is syntax-and-type-checked during editing.
Errors are highlighted in red, and detailed in a text box under the editor.
These informations are refreshed every time the content of the editor changes.
Once the program compiles without error, the Simulate, Verify properties and
Autocorrect tabs shown in \cref{fig:screenshot1} appear.

\paragraph*{Simulate}

The first tab is dedicated to simulation of the program.
By default, it allows the user to set the input values of a selected node, and
displays the output values.
This may be done either manually, as show in \cref{fig:screenshot1}, or by
writing a Lustre expression (e.g. \texttt{0 -> (pre x + 1)} at right of
\texttt{x =}).
Outputs are automatically recomputed if the program is modified
by the user.

The app also supports adding custom simulations for more complex programs where
a longer trace is necessary for simulation. For instance, we wrote a simulator
for a stopwatch based on the seminal example from
Harel~\cite[Fig. 25]{Harel:Statecharts:1987}.
It is shown on \cref{fig:screenshot1}, at right. Clicking on the buttons sends
true signals to the corresponding inputs of the node, and the time displayed is that returned
by the output of the node.
These interactive simulator allows the user to easily test the behavior of their controller.

\paragraph*{Verify properties}

The next tab is only displayed for editors that contain assume-guarantee
contracts.
Clicking it triggers property verification, which may take a few seconds
depending on the size of the program.
When it finishes, the proven properties are highlighted in green in the editor.
Any falsified property is highlighted in red, and a counterexample generated by
the model-checker is displayed.
This interaction is similar to that of the Kind 2 online app~\cite{Kind2App}.

\paragraph*{Autocorrect}

This tab may be disabled in notebooks settings (for instance, during exams).
We say that the students solution is correct when it is functionally equivalent
to the reference solution specified by the teacher (and hidden from the
student).
We use model-checking to check that it is the case.
Clicking the tab sends the student's program to the server, which
model-checks the equivalence between it and the teacher's solution.
If the two do not match, a counterexample is displayed.

\paragraph*{Autosave and Export}

The content of each notebook is automatically saved at each change, which
ensures no work is ever lost. This save is synchronised on any other
computer using the same token.

Moreover, the file menu presented at the top left of the application features two button
to export and import the content of the current notebook.
We use JSON files, which store the content of each editor in a list.
Each file also contains the name of the notebook, which allows us to check
that the correct notebook is being imported. If there is a mistake, (e.g. the
user is trying to import the content of course 1 inside course 2), the import
is cancelled.
This avoids overwriting previous work by mistake.

\subsection{Technical choices and Implementation}

The technical choices we made are motivated by non-functional requirements: we
wanted the simulation to be fast, and to remain so even if several users 
are using the app at the same time.
Therefore, the simulation must run client-side, in the browser, as
we could not afford a powerful server.
To simulate a node efficiently, the application compiles it to
imperative code, following a process we detail below.
Therefore, we needed to embed the Heptagon compiler in the application.
Heptagon happens to be implemented in OCaml~\cite{LeroyEtAl:OCamlMan:2025}, 
a multi-paradigm programming language inspired from
Standard~ML~\cite{MilnerEtAl:DefSML:1997}.
To run it in the application, we used
Js\_of\_ocaml~\cite{Vouillon:JsOfOCaml:2014}, a compiler from OCaml bytecode to
Javascript.

%% TODO figure archi SYNCHRON ?

\paragraph*{Programming Interactions with Js\_of\_ocaml}

We wrote the interactive code for our application in OCaml using Js\_of\_ocaml
bindings to set callbacks and manipulate the \ac{DOM}.
When the content of an editor changes, the application runs the compilation
functions exposed by the Heptagon compiler, and manipulates the \ac{DOM} to
display results (setting error messages, updating simulation results, etc).

\paragraph*{Client-Server Architecture}

To support user authentication and saving notebooks, we developed a simple
server %% in around 300 lines of OCaml 
using the conduit library~\cite{Conduit8} for handling HTTP
requests.
User's notebooks are stored as plain text JSON files compatible with 
those produced by local export.

\paragraph*{Fast Simulation of Lustre Code}

The Heptagon compilation chain includes Obc, a simple object-oriented language
where each Lustre node is represented by a class with attributes representing
the state, a \texttt{reset} method to reinitialise the state, and a
\texttt{step} method to compute one reaction of the node and update the state.
Our first version of the simulator was based on an OCaml interpreter for this
language.
Lustre nodes were compiled into Obc classes, which methods were interpreted on
demand (e.g. when clicking the ``step'' button in \cref{fig:screenshot1}).
However, while this was adequate for simple nodes, the interpreter was too slow
for running some real-time nodes (e.g. a sound-processing node which runs in
real-time in the browser).

To improve the performances, we developped a new Javascript backend for
Heptagon.
Lustre nodes are compiled to Javascript classes which are injected into the
browser's interpreter using the infamous\footnote{\texttt{eval()} may be used to inject malicious code.
We argue that this is not a problem in our case, since (1) our application does
not manipulate any security-critical data and (2) \texttt{eval()} only runs code
generated by the Heptagon compiler, not arbitrary code.}
Javascript \texttt{eval()} function.
Running Javascript code directly means the simulation may profit from the many
optimisations included in the Javascript \ac{JIT} compilers of modern browsers. 

%% TODO example compilation to Js

\paragraph*{Model Checking with Kind~2}

Although Kind~2 is implemented in OCaml, it relies on SMT solvers, such as Z3~\cite{DeMoura:Z3:2008}, which are
often implemented in C for performances reasons.
Therefore, it was not practical to run Kind~2 directly in the browser.
Instead, we run Kind~2 on the same server used for storing user data.
Heptagon programs are translated (client-side) into the syntax of Kind~2, and
sent over the network.
The server processes the request by starting a Kind~2 process, waiting until it
finishes or times out, and sends back the result.

\paragraph*{Creating Notebooks and Exercises}

As of now, there is no way for teacher to create new notebooks through the
application. Instead, notebooks are modified directly in the client application.
A notebook corresponds to a set of \texttt{.html} (for text blocks) and \texttt{.lus} source files, sequenced in a
\texttt{.json} file.
The client application must be recompiled to take the new notebooks into
account.
Similarly, dedicated simulators for specific examples must be implemented in
OCaml following a specified module interface.
These simulators may use additional libraries, such as the OCaml Graphics
library.

The consequence of these restrictions is that, currently, a teacher who would
want to reuse Learn-Heptagon with their own material must deploy their own
instance. They may find the source code for both server and client at \url{https://github.com/learn-heptagon}.

\section{Teaching Lustre Modelling at ENAC}
\label{sec:lesson-plan}

In 2025, we first used Learn-Heptagon to teach a Lustre Modelling course at
\ac{ENAC}, for two groups of students.
In this section, we present the background of the students following this course.
We then present the learning goals of the course, and the learning activies
proposed to the students to achieve these goals.
Finally, we discuss the final assessment, and justify why it covers the learning goal. 

\subsection{Target Audience}

\ac{ENAC} is a French engineering school focused on civil aviation.
It teaches future commercial airline pilots, air traffic controllers, and engineers
specialized in air traffic control systems and avionics.
Our course on Lustre-based modelling is offered to students in two
curricula, both focused on air traffic systems engineering.

\begin{itemize}
%% \item \ac{IENAC} is a three-year degree. Students are recruted through an
%% entrance exam after 2 years of \ac{CPGE} which gives them general training in
%% maths, physics and computer science.
%% The first year presents the basics of aviation and air traffic management.
%% In the second year, they choose one of four specialization, one of which is
%% focused on computer science.
%% In the third year, the students having chosen this specialisation may take the
%% \ac{ISI} option which specializes into engineering of interactive software; our
%% course is offered only in this option. This represents less than 5 students
%% every year, with the option not opening in the last 2 years because of a lack of students.
\item \ac{IENAC APPR} is a three years apprencticeship-based engineering degrees: students
  follow some general engineering course, as well as courses specialised in
  aviation and air traffic management. They spend half their
  time working in a company to acquire more professional skills.
  Students usually come from practical background, some already having a
  professional experience. %% , and are selected on their CV.
  In the last year, students may take the \ac{SI} option which is focused on
  computer science and includes our course. This represents around 15 students every year.
\item \ac{IATSED} is a two years specialized master's degree. Students come from
  multiple countries and have varied backgrounds, usually having followed an
  undergraduate curriculum. %% , and are selected on their CV.
  Our course is mandatory in their second year. This represents 10 to 15 students every year.
\end{itemize}

Since our course is offered in the last year of each curriculum, we considered 
the courses related to modelling and verification offered at \ac{ENAC} overall.
Both curricula include an early course on requirement engineering, during which students learn
how to find system requirements and express them in natural language.
Then, there is a \ac{VV} course %% usually offered in the second year,
which focuses on how requirements are validated when designing of the system,
mostly with test-based methodologies.
The second part of the course discusses software testing more specifically.
Finally, \ac{IATSED} students are offered three courses on modelling
and verification using formal methods in their last year: our Lustre modelling
course, a course based on
Event-B~\cite{Abrial:EventB:2010}, and a course on abstract interpretation.
\ac{IENAC APPR} students are only offered the Lustre modelling course.

Notably, no curriculum discussed above includes a course focused on logic.
Some of the \ac{IATSED} students may have acquired some knowledge during their
undergrad-level courses, but it is not required for recruitement.
%% Some of the computer science courses proposed to the \ac{IENAC} may also include some
%% applications of logic (e.g. relational databases, operations research), but
%% there is no theoretical course on logic.
\ac{IENAC APPR} also often lack theoretical bases, due to their more practical
background, and are not offered any theoretical course on logic either, and are
often not very familiar with logical reasoning (e.g. reasoning by contradiction
or by induction).
One major challenge of the course is therefore to work around this lack of
pre-requisite knowledge.

%% TODO rappeller que ce n'est pas des prog

%% Cours de V&V: d'abord général, puis plus orienté tests avec focus software

%% IATSED:
%% V&V -> S7
%% Needs Engineering -> FM V&V 1 -> FM V&V 2
%% Crédits: 4 + 6 + 9 = 19 / 90
%% No math/logic course

%% IENAC SITA:
%% Logique en prépa mais pas dans la formation ?
%% Verif et Validation -> S8
%% Systemes adaptatifs, FM V&V

%% IENAC APPR:
%% stage remise a niveau ?
%% UE Maths/Info : 9 + 10 + 5 + 5
%% V&V en S8

\subsection{Learning Objectives}

As discussed above, the students undertaking this course are not 
computer scientist, but rather system engineers.
Therefore, we approach the course from the point of view of software
engineering.
We use synchronous programming, and Lustre in particular as tools to model
embedded systems.
Therefore, at the end of the course, we expect students to be able to 
\textbf{model an embedded system and its properties as a Lustre program}.

Since our students are usually not passionate about programming languages, one
intermediate objective of the course is to convince them of the relevance of
synchronous programming for safety-critical embedded systems.
We expect the students to be able to
\textbf{explain the advantages of the synchronous model over other paradigms for
  embedded systems},
and to \textbf{explain the constraints imposed by the synchronous model}.

To reach the general objective of the course, the course first need to tackle two
intermediate objectives.
First, the students must be able to
\textbf{write synchronous programs using the core Lustre syntax}.
A first iteration of the course tried to also tackle
the main extensions of Scade~6 (arrays and state machines), but we
realised that this was too ambitious, given the time allocated for the course
(8 hours) and the background of our students.
Finally, students must be able to
\textbf{specify the requirements of a system as assume-guarantee contracts}.

%% TODO Taxonomy levels ? relatively high (focus on second half of the objectives)

\subsection{Organisation and Learning Activities}

The course is divided into two roughly equivalent parts: first, introducing
synchronous programming and the Lustre language, then focusing on specification
with assume-guarantee contracts.
We detail the content of each part below.%
\footnote{
The slides are provided as supplementary material with this submission, and the
exercises are available on \url{https://learn-heptagon.vertmo.org/} (notebooks
``Course~1: Programming in Lustre'', and ``Course~4: Contracts and Verification'').}

\paragraph*{Organisation}
The course is given over two to three weeks in four sessions of two hours each.
%% , usually taught  depending on the agenda and teaching load of the students.
Each session contains a mix of lecture and practical work (using
Learn-Heptagon), following the lesson plan outlined below.
All exercises are done synchronously in class.
During this practical work, we help the students and answer their question.
We typically give the exercises by batch, and wait until around three quarters
of the students have finished the batch.
Then, we correct the exercises on the board.

\paragraph*{Part 1: Introduction to synchronous dataflow programming in Lustre}
Part of this course was adapted from lectures by C. Tinelli at The University of
Iowa, themselves adapted from lectures by P. Raymond at Université Grenoble
Alpes. It also includes exercises designed by T. Bourke for his lectures at
Sorbonne Université.

The start of the course aims at introducing the synchronous dataflow paradigm.
To do so, it highlights the usual design of an embedded controller, which samples
its environment at a given rate, computes and sends orders to actuators.
It justifies the need for dedicated languages that are closer than C to
high-level system specifications.
It also justifies why abstracting physical time simplifies programming such systems,
and introduces the synchronous hypothesis and the family of synchronous languages.
Finally, it introduces synchronous dataflow languages, and Lustre in particular,
first through simple block-diagram circuit, which are shown to translate to
Lustre node; this also mirrors the design of SCADE.
This first theoretical part of the course is given as a 30 minutes lecture.

Then, the first practical part of the course focuses on combinatorial programs
that do not manipulate a state (e.g. that do not 
contain a \lus{->} or \lus{pre}).
These programs are easier to understand, since each reaction is
independent from the previous ones.
The syntax of the few Lustre constructs is introduced (node declaration and
call, equations, arithmetic and logic operators).
We focus in particular on the \lus{if-then-else} construct: since our students are
not used to functional programming, \lus{if} being an expression is often
confusing for them.
With that in mind, we ask the students to write their first few simple
combinatorial nodes which involve \lus{if} (calculating the max of two inputs,
the absolute value of an input, etc). 
The second exercise is slightly more involved: it requires students to implement
a 1-bit binary
adder~\footnote{\url{https://en.wikipedia.org/wiki/Adder_(electronics)}} using
logical gates (xor, and, or).
A first version must be defined as a single node, which uses seven logical gates.
For the second version, students must first define a node that implements a half
adder, and compose two calls to this node to define the full adder.
As support, the students are given the truth table of the full adder, and
schematics of the expected circuits.

The second practical part is focused on the stateful operators: \lus{->} and
\lus{pre}.
When first introduced, we insist on the problems that they may cause if used
incorrectly (\lus{pre} not being defined at the initial step, possibility of
introducing causality loops).
Then, we ask the students to program a few simple stateful nodes (compute the
cumulative sum of inputs, flip/flop gate, etc). 

This part of the course is capped by two more realistic exercises, both using a
custom simulator in Learn-Heptagon.
The first consists in programming a running average over a number of samples of a noised signal
to smooth it.
The second consists in implementing the controller for the stopwatch shown in
\cref{fig:screenshot1}.
%% which ticks every
%% hundredth of a second, with buttons to start and stop, reset and freeze the
%% display.
Students typically struggle with this final exercise, but they are given enough 
time and guidance so that about half of the class manages to finish it.
These examples could be used, in a more comprehensive course, to illustrate the
expressivity of arrays for writing the running average and state machines for
writing the stopwatch controller.

\paragraph*{Part 2: Specifying and Verifying Program Properties}

The second part of the course starts with a quick introduction of the
compilation scheme of Lustre programs into imperative C code.
This is mainly intended as a reminder that these high-level dataflow synchronous
programs are indeed compiled efficiently.

Then, the course discusses formal verification of Lustre
programs.
It starts with a general introduction of Hoare
triples~\cite{Hoare:Triples:1969}, which, in this context, relates to the notion of synchronous
observers.
One advantage of synchronous observers is that they do not require additional syntax,
and are easy to understand by students at this point.
We introduce the idea of model-checking as a way to check a property
expressed by a synchronous observer exhaustively; the inner working of model
checkers are out of the scope of this course.

Then, the course builds on this by introducing
\luskw{assume}/\luskw{guarantee} contracts. As a break, students write a
few simple logical nodes (such as the \texttt{sofar} node from the
introduction), which are then used in a more complete example: the specification
of the flip/flop node from the previous practical work.
In particular, the example shows that the (conditional) equivalence between two possible
implementations of the flip/flop may be proven through model-checking.

This part of the course is capped by three modelling exercises for more
complex systems.
First, a guided exercise of specifying properties of a traffic light.
The implementation is provided, and students need to add a contract encoding a
set of given requirements, which are intended to be precise and easy to
formalise (e.g. ``The pedestrian green light is only on when the car red light is'').
The second exercise is a bit more abstract, and still has a precise contract,
but also asks the student to implement the node.
The final exercise is more open-ended: students are asked to implement the
controller for a robot that follows the voice of its owner.
The controller is relatively high level (outputs only indicate if the robot
should rotate or move forward).
The specification is intentionally left vague, aside from the requirements, which
guides the students towards first specifying the system before starting
implementing the controller.

%% \begin{itemize}
%% \item Course 1
%%   \begin{itemize}
%%   \item intro to synchronous programming (~fast: 15min). maybe it could be made longer ?
%%   \item TODO
%%   \item more realistic exercises: sliding average + state machines. Each may be revisited later (courses 3 and 4)
%%   \end{itemize}
%% \item Course 2
%%   \begin{itemize}
%%   \item compilation
%%   \item clocks
%%   \item state machines
%%   \end{itemize}
%% \item Course 3
%%   \begin{itemize}
%%   \item arrays. previous courses: different version (Lustre v4, v6, Heptagon). now focusing on Heptagon
%%   \item exercise: fifo used for to write audio filters (FIR and IIR = reverb)
%%   \item array iterators.
%%   \item iterator exercise: sliding average 
%%   \end{itemize}
%% \item Course 4
%%   \begin{itemize}
%%   \item hoare logic
%%   \item synchronous observers (avantage: pas de nouvelle syntaxe)
%%   \item assume-guarantee contracts, with the syntax
%%   \item program some logical nodes (maybe this could be before ?)
%%   \item a given example (switch node seen in course 1)
%%   \item exercises: traffic light, cycles, robot
%%     @PLOC: why in that order ? What was the reasoning for these specific exercises ?
%%   \end{itemize}
%% \end{itemize}

\subsection{Final Assessment}

At the end of the course, we assess what the students have learned through a
2 hours computer-based exam.
The exam questions are similar to the exercises practiced during the second part
of the course.
Students are asked to use assume-guarantee contracts to specify a system
described on paper, and to implement a controller that accomplishes this
specification.
We choose realistic yet simple embedded systems; in particular, we do not want
the students to struggle with complex control laws.
One recurring theme has been monitoring systems which check the
outputs of another more complex system, and go into an error mode if these
outputs do not respect some constraints; this has been applied, in particular,
to a drone controller and to a battery monitoring system.
We typically split the subject into two to three independent exercises to limit
the risk of students getting stuck on a question.

In 2025, we used Learn-Heptagon during the exam without any issue.
During the exam, students have access to the simulator and property verification
features of learn-heptagon, but not to the autocorrect feature.
At the end of the exam, we ask the students to export their work and send the
resulting files to us. As a backup, we can also access their work on the server,
or connect using their token if necessary.

\section{Conclusion}

\subsection{Observations and Feedback}

Although our students were not computer science or logic experts, they did take
to the tools and to the proposed exercise without too many issues.
We believe that this is due to the mix of lectures and practical sessions in a
short period, which allows the students to start working first on simple subsets
of the language, progressing to more complex features of the language.
In previous iterations of the course, during which lectures and practical sessions were
more separated, student often had trouble starting their first Lustre code,
possibly intimidated by a new syntax.
The Learn-Heptagon application is particularly adapted to this workflow, as
students did not waste time setting up tools and environments when starting a
batch of exercises. 

One issue with mixing lectures and exercises during a lesson is that all
students do not progress at the same speed, leading to some waiting for others
to finish a batch of exercises. 
One solution to this problem is to provide ``bonus'', more complex exercises for
fast students.
While we did not plan for it for the 2025 iteration of the course, the application
would allow for incorporating such exercises in the notebook, possibly with some
custom styling.

One recurring pain point for the student was writing programs that use floating
point numbers.
Indeed, in Heptagon, floating point operators must be written with a
trailing~\texttt{.} : for instance, the expression \lus{2. +. 1. <. 3.} is correctly
typed.
Students often forget these~\texttt{.}, which leads to typing issues. Although
these are clearly highlighted in our application, the error message is not clear
enough for students not used to statically-typed languages.

The student did evaluate the course using \ac{ENAC}'s feedback tool.
No comment was directed specifically at Learn-Heptagon.
One trend in the comments was that a number of \ac{IENAC APPR} students did not understand
the purpose of learning the Lustre language, as they found it ``too niche''.
We believe that we were not explicit enough in explaining that the objective of
the course was learning to formally model a safety-critical embedded
systems, and that Lustre was a useful tool to do so.
We will dedicate more time to motivating the course in future iterations.

During the course, we detected a few bugs in our application, which have been
fixed since then.
There is a more problematic design issue in the autocorrect feature.
Checking the equivalence of the student's node with the teacher's solution may be too
restrictive when the behavior of the node is not fully defined by the
specification.
One way to alleviate this issue could be to check the equivalence under some
conditions provided with the teacher's solution.
Finally, we found some bugs in the Heptagon compiler, mostly related to the
compilation of programs using arrays. They have been reported to the developer
of Heptagon.

\subsection{Discussion and Future Work}

The design choices of our application are inspired from existing web-based
applications for teaching programming languages.
The similarly-named Learn~OCaml~\cite{Canou:LearnOcaml:2016} is dedicated to the
OCaml multi-paradigm language.
It allows the teacher to define exercises by writing their specification in
Markdown files and defining a code canvas to be filled by the students.
It also allows to automatically test the answers of the students: the teacher
writes a correction, and a set of test cases to compare the teacher's and
student's solutions on.
One Learn~OCaml feature absent from Learn-Heptagon is the ``Teacher'' page,
which allows the teacher to access the answers of students, and gives detailed
statistics about the completion rates of the different exercises.
Such a feature would be useful in Learn-Heptagon.
To implement it, we would need to save the completion status of
each exercise (not attempted, does not compile, compiles but autocorrect fails,
succeeds).
This would also allow to present a bird's eye view of their progress to
students.

The notebook presentation of exercises is not inspired from Learn OCaml, but
from Jupyter Notebook, an interactive programming environment initially
developped for Python.
Jupyter Notebooks have been used extensively for teaching programming and other
scientific subjects~\cite{Ochkov:JupyterSTEM:2022}.
Classic Jupyter notebooks rely on a local client-server architecture where
kernels are loaded in the server for the different languages.
The web page displayed in the browser interacts with this server.
There is a server-less version called JupyterLite which runs entirely in the
browser, but is more limited in the choice of languages, since JupyterLite
kernels must also run in the browser.
It would be possible to export the simulator component of Learn-Heptagon as a
JupyterLite kernel, but not the property verification component as it is, since
it runs on the server.
It would also be possible to integrate all features of Learn-Heptagon into a
Jupyter Notebook kernel.
This would not however solve the issue of installing this environment on student
machines.
%% TODO dynamic notebook editing ?

In the current iteration of our course, all learning activities are individual.
We may want to include more ambitious group projects at the end of the course.
To support such a group activity, Learn-Heptagon would need to integrate tools
for collaborative work.
At its most basic, this could mean integration with a \ac{VCS}, like Git.
A more interactive approach would be to integrate real-time collaborative editing
in the code editors.
We could either integrate an editor that already supports this feature, such as
MUTE~\cite{Nicolas:MUTE:2017}, or extends the one we already use with
synchronisation algorithms based on \acp{CRDT}~\cite{Shapiro:CRDT:2011}.

Our translation from the syntax of Heptagon to that of Kind 2 is currently
incomplete.
Indeed, Kind 2 does not support state machines, and only supports a limited form
of arrays, while Heptagon supports a more general form of these features.
This means that model-checking, and therefore autocorrect, cannot be used on
programs that use these features.
It should be possible to translate both of these features to the syntax of Kind 2.
First, since the size of arrays is known at compile time, it is possible to expand
any array of size $n$ into $n$ independent variables; this was the approach of
Lustre v4~\cite[sec 3.1]{HalbwachsRay:Lustrev4Tutorial:2007}.
This does not produce efficient code, but should enable model-checking for
arrays small enough.
Second, Heptagon translates state machines into its kernel language using the \lus{when}
and \lus{merge} operators to sample and recombine sampled streams
respectively~\cite{ColacoPagPou:StateMachines:2005}.
Kind 2 only supports a very limited form of these sampling operators, which is
in general not compatible with the code generated from a state machines.
We are currently experimenting with a translation pass that transforms a sampled
program into an equivalent unsampled program.
Applying this pass before the translation to Kind 2 should enable model-checking 
of programs with state machines.

Over the past few years, programmers have started using \ac{GenAI}, and
\acp{LLM} in particular, to
automate writing code.
However, studies~\cite{Lepp:GenAI:2025} have shown that using \ac{GenAI} tools in the
academic setting negatively correlates with performances, at least in the short
term.
Since all exercises sessions in our course are done on the computer, it is tempting
for our students to use an \ac{LLM} to generate a solution rather than searching
for it themselves.
Fortunately, since Lustre is a relatively uncommon language, most \acp{LLM} do
not seem to have enough training data to be proficient with it.
This is not however always enough to discourage students from using these tools.
Therefore, we are still considering ways to nudge them away from \ac{GenAI}.
In addition, we formally forbid students from using \acp{LLM} during the exam.
We closely monitor students to ensure they do not use it, but this is only
possible because we have small classes.
In the future, we may need to integrate automated monitoring and protections
against \ac{LLM}-based cheating.
For instance, we could detect large pastes inside an editor.
We could also use a tool similar to SafeExamBrowser~\cite{SafeExamBrowser} to prevent the students from
leaving Learn-Heptagon's page during the duration of the exam.

%% TODO when do we discuss about how happy the students where about the course ?
%% Conclusion or in sec 3 ?

%%
%% Bibliography
%%

\bibliography{bib/abbrevs-short,bib/abbrevs,bib/refs}

@string{ieee =		"{IEEE}"}

@string{workshop =	"Workshop on"}

@string{lncs =		"LNCS"}

@string{cup =		"{Cambridge University Press}"}

@string{lncs =		"Lecture Notes in Elecrical Engineering"}

@string{Revised =	"Revised"}

@string{and =		"and"}

@string{Software =	"Software"}

@string{Engineering =	"Engineering"}

@string{cup =		"{CUP}"}

@string{Revised =	"Rev."}

@string{and =		"\&"}

@string{Software =	"Soft."}

@string{Engineering =	"Eng."}

@string{acm =		"{ACM}"}

@string{inria =		"Inria"}

@string{acmpress =	"{ACM Press}"}

@string{mitpress =	"{MIT Press}"}

@string{springer =	"{Springer-Verlag}"}

@string{springer =	"{Springer}"}

@string{procieee =	proc # " " # ieee}

@string{cav16b =	proc # " 28th " # cav # "2016), Part {II}"}

@string{date =		"Design, Automation and Test in Europe (DATE"}

@string{emsoft =	intcon # " Embedded Software (EMSOFT"}

@string{emsoft05 =	proc # " 5th " # acm # " " # emsoft # " 2005)"}

@string{fmcad00 =	proc # " 3rd " # intcon # " " # fmcad # "2000)"}

@string{fmcad08 =	proc # " 8th " # intcon # " " # fmcad # "2008)"}

@string{lctes =		acm # " " # sigplan # " " # conf #
	    " Languages, Compilers, and Tools for Embedded Systems (LCTES"}

@string{lctes08 =	proc # " 9th " # lctes # " 2008)"}

@string{popl87 =	proc # " 14th " # popl # " 1987)"}

@string{programming =	intsym # " Programming"}

@string{sac15 =		proc # " 30th " # sac # "'15)"}

@string{sat =		intcon # " Theory and Applications of Satisfiability
			Testing (SAT "}

@string{vmcai11 =	proc # " 12th " # vmcai # " 2011)"}

@inproceedings{Papailiopoulou:Coverage:2011,
  title = {Structural {{Test Coverage Criteria}} for {{Integration Testing}} of {{LUSTRE}}/{{SCADE Programs}}},
  booktitle = {Formal {{Methods}} for {{Industrial Critical Systems}}},
  author = {Papailiopoulou, Virginia and Rajan, Ajitha and Parissis, Ioannis},
  editor = {Salaün, Gwen and Schätz, Bernhard},
  date = {2011},
  pages = {85--101},
  publisher = {Springer},
  location = {Berlin, Heidelberg},
  doi = {10.1007/978-3-642-24431-5_8},
  isbn = {978-3-642-24431-5},
  langid = {english},
}

@article{Harel:Statecharts:1987,
  author =	{David Harel},
  title =	{Statecharts: A Visual Formalism for Complex Systems},
  journal =	{Science of Computer Programming},
  volume =	8,
  number =	3,
  month =	jun,
  pages =	{231--274},
  year =	1987,
  review =	{20030915},
  url =		{http://www.inf.ed.ac.uk/teaching/courses/seoc/2005_2006/resources/statecharts.pdf},
  doi =		{10.1016/0167-6423(87)90035-9},
}

@article{BenvenisteBer:SyncApp:1991,
  author =	{Albert Benveniste and G\'{e}rard Berry},
  title =	{The Synchronous Approach to Reactive and Real-Time Systems},
  journal =	procieee,
  year =	1991,
  VOLUME=	79,
  number =	9,
  pages =	{1270--1282},
  month =	sep,
  organization =ieee,
  url =		{https://ptolemy.berkeley.edu/projects/chess/design/2010/discussions/Pdf/synclang.pdf},
  review =	{20031117},
}

@article{Hoare:Triples:1969,
  author    = {C. A. R. Hoare},
  title     = {An Axiomatic Basis for Computer Programming},
  journal   = {Commun. ACM},
  volume    = {12},
  number    = {10},
  year      = {1969},
  pages     = {576-580},
  ee        = {http://doi.acm.org/10.1145/363235.363259},
  bibsource = {DBLP, http://dblp.uni-trier.de}
  }

@inproceedings{CaspiEtAl:Lustre:1987,
  author =      {Paul Caspi
		 and Daniel Pilaud
		 and Nicolas Halbwachs
		 and Plaice, John A.},
  title =	{{LUSTRE}: A declarative language for programming
		 synchronous systems},
  crossref =	{POPL1987},
  pages =	{178--188},
  url = {https://www.cse.unsw.edu.au/~plaice/archive/JAP/P-ACM_POPL87-lustre.pdf},
  doi =		{10.1145/41625.41641},
}

@manual{JahierRayHal:Lustre6:2019,
  title =	{The {Lustre V6} Reference Manual},
  author =	{Erwan Jahier
		 and Pascal Raymond
		 and Nicolas Halbwachs},
  organization ={Verimag},
  address =	{Grenoble},
  year =	2019,
  month =	may,
  url =
  {http://www-verimag.imag.fr/DIST-TOOLS/SYNCHRONE/lustre-v6/doc/lv6-ref-man.pdf},
}

@misc{Raymond:Lustre4:1992,
  author =	{Pascal Raymond},
  title =	{The {Lustre V4} Distribution},
  howpublished ={\url{http://www-verimag.imag.fr/The-Lustre-Toolbox.html}},
  year =	1992,
  month =	sep,
}

@manual{HalbwachsRay:Lustrev4Tutorial:2007,
  title =	{A Tutorial of {Lustre}},
  author =	{Nicolas Halbwachs and
		 Pascal Raymond},
  organization =     {Verimag},
  address =	{Gières, France},
  year =	2007,
  month =	aug,
  url =		{http://www-verimag.imag.fr/DIST-TOOLS/SYNCHRONE/lustre-v4/distrib/lustre_tutorial.pdf},
}

@manual{Heptagon-Tutorial:2017,
  title =	{Heptagon/{BZR} manual},
  author =	{Heptagon Developers},
  year =	2017,
  month =	apr,
  url =		{http://heptagon.gforge.inria.fr/pub/heptagon-manual.pdf},
}

@inproceedings{HagenTin:Kind:2008,
  author =	{George Hagen and Cesare Tinelli},
  title =	{Scaling Up the Formal Verification of {Lustre} Programs
		 with {SMT}-based Techniques},
  crossref =	{FMCAD08},
  pages =	{Article 15},
  url =		{http://homepage.cs.uiowa.edu/~tinelli/papers/HagTin-FMCAD-08.pdf},
}

@inproceedings{ChampionEtAl:Kind2:2016,
  author =	{Adrien Champion
		 and Alain Mebsout
		 and Christoph Sticksel
		 and Cesare Tinelli},
  title =	{The {Kind 2} Model Checker},
  crossref =	{CAV16b},
  pages =	{510--517},
  doi = {10.1007/978-3-319-41540-6_29}
}

@misc{Ansys:Scade,
  shorthand =	{SS},
  author =	{{ANSYS/Esterel Technologies}},
  title =	{{SCADE Suite}},
  howpublished ={\url{http://www.ansys.com/products/embedded-software/ansys-scade-suite}},
}

@manual{Scade6Language:2016,
  shorthand =	{S6L},
  title =	{The {SCADE~6} Language},
  author =	{{ANSYS/Esterel Technologies}},
  year =	2016,
  month =	mar,
  note =	{KCG-SRS-007/R/2},
}

@inproceedings{BiernackiEtAl:ClockDirected:2008,
  author =	{Dariusz Biernacki and
		 Jean-Louis Cola{\c{c}}o and
		 Gregoire Hamon and
		 Marc Pouzet},
  title =	{Clock-directed modular code generation for synchronous
		 data-flow languages},
  crossref =	{LCTES08},
  pages =	{121--130},
  url =		{https://www.di.ens.fr/~pouzet/bib/lctes08a.pdf},
}

@inproceedings{DieumegardEtAl:SyncObs:2015,
  author =	{Arnaud Dieumegard
		 and Pierre-Lo{\"{i}}c Garoche
		 and Temesghen Kahsai
		 and Alice Taillar
		 and Xavier Thirioux},
  title =	{Compilation of Synchronous Observers as Code Contracts},
  booktitle =	sac15,
  year =	2015,
  month =	apr,
  pages =	{1933--1939},
  address =	{Salamanca, Spain},
  publisher =   acmpress,
}

@inproceedings{ColacoPagPou:StateMachines:2005,
  author =	{Jean-Louis Cola{\c{c}}o and Bruno Pagano and Marc Pouzet},
  title =	{A Conservative Extension of Synchronous Data-flow with
		 State Machines},
  crossref =	{EMSOFT05},
  pages =	{173--182},
  url =		{https://www.di.ens.fr/~pouzet/bib/emsoft05b.pdf},
  doi =		{10.1145/1086228.1086261},
  review =	{20110117},
}

@book{Abrial:EventB:2010,
  author =	{Jean-Raymond Abrial},
  title =	{Modeling in {Event-B}: System and Software Engineering},
  publisher =	cup,
  year =	2010,
  month =	jun,
  url =		{http://www.event-b.org/abook.html},
}

@incollection{Lichtenstein:PLTL:1985,
  title = {The Glory of the Past},
  booktitle = {Logics of {{Programs}}},
  author = {Lichtenstein, Orna and Pnueli, Amir and Zuck, Lenore},
  editor = {Parikh, Rohit},
  editora = {Goos, G. and Hartmanis, J. and Barstow, D. and Brauer, W. and Brinch Hansen, P. and Gries, D. and Luckham, D. and Moler, C. and Pnueli, A. and Seegmüller, G. and Stoer, J. and Wirth, N.},
  editoratype = {redactor},
  date = {1985},
  volume = {193},
  pages = {196--218},
  publisher = {Springer Berlin Heidelberg},
  location = {Berlin, Heidelberg},
  doi = {10.1007/3-540-15648-8_16},
  url = {http://link.springer.com/10.1007/3-540-15648-8_16},
  urldate = {2026-04-15},
  isbn = {978-3-540-15648-2 978-3-540-39527-0},
  langid = {english}
}

@inproceedings{SheeranSinSta:kInduction:2000,
  author =	{Mary Sheeran
		 and Satnam Singh
		 and Gunnar St{\aa}lmarck},
  title =	{Checking Safety Properties Using Induction and a
		 {SAT}-Solver},
  crossref =	{FMCAD00},
  pages =	{127--144},
  url =		{https://www.di.ens.fr/~pouzet/cours/mpri/sheeran-FMCAD00.pdf},
}

@inproceedings{Bradley:IC3:2011,
  author =	{Bradley, Aaron R.},
  title =	{{SAT}-Based Model Checking without Unrolling},
  crossref =	{VMCAI11},
  pages =	{70--87},
}

@book{LeroyEtAl:OCamlMan:2025,
  author =	{Xavier Leroy and Damien Doligez and Alain Frisch
		 and Jacques Garrigue and Didier R{\'{e}}my
         and KC Sivaramakrishnan
		 and J{\'{e}}r{\^{o}}me Vouillon},
  title =	{The {OCaml} system: Documentation and user's manual},
  publisher =	{Inria},
  year =	2025,
  edition =	{5.4},
  month =	oct,
}

@book{MilnerEtAl:DefSML:1997,
  author =	{Robin Milner and Mads Tofte and Robert Harper
		 and David MacQueen},
  title =	{The Definition of {Standard ML} (Revised)},
  publisher =	mitpress,
  year =	1997,
  month =	may,
}

@manual{DO-331,
  shorthand =	{DO-331},
  title =	{Model-Based Development and Verification Supplement
		 to {DO-178C} and {DO-278A}},
  organization ={RTCA, Inc.},
  address =	{Washington, DC, USA},
  year =	2011,
  month =	dec,
  url =		{https://standards.globalspec.com/std/1460383/rtca-do-331},
}

@proceedings{CAV16b,
  title =	cav16b,
  booktitle =	cav16b,
  year =	2016,
  editor =	{Swarat Chaudhuri and Azadeh Farzan},
  series =	lncs,
  volume =	9780,
  address =	{Toronto, Canada},
  month =	jul,
  publisher =	springer,
}

@proceedings{EMSOFT05,
  key =		{EMSOFT},
  title =	emsoft05,
  booktitle =	emsoft05,
  year =	2005,
  editor =	{Wayne Wolf},
  address =	{Jersey City, USA},
  month =	sep,
  publisher =	acmpress,
}

@proceedings{FMCAD00,
  title =	fmcad00,
  booktitle =	fmcad00,
  year =	2000,
  editor =	{{Hunt Jr.}, Warren A.
		 and Johnson, Steven D.},
  address =	{Austin, TX, USA},
  month =	nov,
  organization =ieee,
}

@proceedings{FMCAD08,
  title =	fmcad08,
  booktitle =	fmcad08,
  year =	2008,
  editor =	{Alessandro Cimatti and Jones, Robert B.},
  address =	{Portland, OR, USA},
  month =	nov,
  organization =ieee,
}

@proceedings{LCTES08,
  key =		{LCTES},
  title =	lctes08,
  booktitle =	lctes08,
  year =	2008,
  address =	{Tucson, AZ, USA},
  month =	jun,
  publisher =	acmpress,
}

@proceedings{POPL1987,
  title =	popl87,
  booktitle =	popl87,
  year =	1987,
  address =	{Munich, Germany},
  month =	jan,
  publisher =	acmpress,
}

@proceedings{VMCAI11,
  title =	vmcai11,
  booktitle =	vmcai11,
  year =	2011,
  month =	jan,
  editor =	{Ranjit Jhala
		 and David Schmidt},
  series =	lncs,
  volume =	6538,
  address =	{Austin, TX, USA},
  publisher =	springer,
}

@article{Vouillon:JsOfOCaml:2014,
  title = {From Bytecode to {{JavaScript}}: The {{Js}}\_of\_ocaml Compiler},
  shorttitle = {From Bytecode to {{JavaScript}}},
  author = {Vouillon, Jérôme and Balat, Vincent},
  date = {2014-08},
  journaltitle = {Software: Practice and Experience},
  shortjournal = {Softw Pract Exp},
  volume = {44},
  number = {8},
  pages = {951--972},
  issn = {0038-0644, 1097-024X},
  doi = {10.1002/spe.2187},
  url = {https://onlinelibrary.wiley.com/doi/10.1002/spe.2187},
  urldate = {2024-12-04},
  langid = {english},
}

@inproceedings{Canou:LearnOcaml:2016,
  TITLE = {{Learn OCaml, An Online Learning Center for OCaml}},
  AUTHOR = {Canou, Benjamin and Henry, Gr{\'e}goire and Bozman, {\c C}agdas and Le Fessant, Fabrice},
  URL = {https://inria.hal.science/hal-01352015},
  BOOKTITLE = {{OCaml Users and Developers Workshop 2016}},
  ADDRESS = {Nara, Japan},
  YEAR = {2016},
  MONTH = Sep,
  HAL_ID = {hal-01352015},
  HAL_VERSION = {v1},
}

@electronic{Garoche:lustrec:2012,
 author = {Garoche, Pierre-Loïc and Thiroux, Xavier and Kahsai, Temesghen},
 note = {collaboration avec l'IRIT et NASA Ames},
 title = {LustreC: a modular Lustre compiler},
 url = {https://github.com/coco-team/lustrec},
 year = {2012}
}

@INPROCEEDINGS{Ochkov:JupyterSTEM:2022,
  author={Ochkov, Valery F. and Stevens, Alan and Tikhonov, Anton I.},
  booktitle={2022 VI International Conference on Information Technologies in Engineering Education (Inforino)}, 
  title={Jupyter Notebook, JupyterLab – Integrated Environment for STEM Education}, 
  year={2022},
  volume={},
  number={},
  pages={1-5},
  doi={10.1109/Inforino53888.2022.9782924}}

@misc{JupyterNotebook,
  title={{The Jupyter Notebook}},
  author={Team, Jupyter},
  url={https://jupyter-notebook.readthedocs.io/en/stable/notebook.html},
  year={2015}
}

@misc{AceEditor,
title={{Ajax.org Cloud9 Editor}},
author={Ajax.org},
url={https://github.com/ajaxorg/ace},
year={2022}}

@misc{Kind2App,
  author =	{Adrien Champion and Alain Mebsout and Christoph Sticksel and Cesare Tinelli},
  title ={Kind 2 Web Application},
  url = {https://kind.cs.uiowa.edu/app/},
  year={2016}
}

@inproceedings{DeMoura:Z3:2008,
  title = {Z3: {{An Efficient SMT Solver}}},
  shorttitle = {Z3},
  booktitle = {Tools and {{Algorithms}} for the {{Construction}} and {{Analysis}} of {{Systems}}},
  author = {de Moura, Leonardo and Bjørner, Nikolaj},
  editor = {Ramakrishnan, C. R. and Rehof, Jakob},
  date = {2008},
  pages = {337--340},
  publisher = {Springer},
  location = {Berlin, Heidelberg},
  doi = {10.1007/978-3-540-78800-3_24},
  isbn = {978-3-540-78800-3},
  langid = {english}
}

@misc{Conduit8,
  author={Anil Madhavapeddy and Thomas Leonard and Thomas Gazagnaire and Rudi Grinberg},
  title={{conduit 8.0.0}},
  year={2025},
  url={https://opam.ocaml.org/packages/conduit/}
}

@inproceedings{Shapiro:CRDT:2011,
  title = {{Conflict-Free Replicated Data Types}},
  booktitle = {Stabilization, {{Safety}}, and {{Security}} of {{Distributed Systems}}},
  author = {Shapiro, Marc and Preguiça, Nuno and Baquero, Carlos and Zawirski, Marek},
  editor = {Défago, Xavier and Petit, Franck and Villain, Vincent},
  date = {2011},
  pages = {386--400},
  publisher = {Springer},
  location = {Berlin, Heidelberg},
  doi = {10.1007/978-3-642-24550-3_29},
  isbn = {978-3-642-24550-3},
  langid = {english}
}

@inproceedings{Nicolas:MUTE:2017,
  title = {{{MUTE}}: {{A Peer-to-Peer Web-based Real-time Collaborative Editor}}},
  shorttitle = {{{MUTE}}},
  booktitle = {Reports of the {{European Society}} for {{Socially Embedded Technologies}}},
  author = {Nicolas, Matthieu and Elvinger, Victorien and Oster, Gérald and Ignat, Claudia-Lavinia and Charoy, François},
  date = {2017-08},
  series = {Proceedings of 15th {{European Conference}} on {{Computer-Supported Cooperative Work}} - {{Panels}}, {{Posters}} and {{Demos}}},
  volume = {1},
  number = {3},
  pages = {1--4},
  publisher = {EUSSET},
  location = {Sheffield, United Kingdom},
  doi = {10.18420/ecscw2017_p5},
  url = {https://inria.hal.science/hal-01655438},
  urldate = {2026-04-27}
}

@article{Lepp:GenAI:2025,
title = {Does generative AI help in learning programming: Students’ perceptions, reported use and relation to performance},
journal = {Computers in Human Behavior Reports},
volume = {18},
pages = {100642},
year = {2025},
issn = {2451-9588},
doi = {https://doi.org/10.1016/j.chbr.2025.100642},
url = {https://www.sciencedirect.com/science/article/pii/S2451958825000570},
author = {Marina Lepp and Joosep Kaimre}
}

@misc{SafeExamBrowser,
title={{SafeExamBrowser}},
author={Information Technology Services unit, ETH Zürich},
url={https://github.com/SafeExamBrowser},
year={2026}}

\appendix

\end{document}